\documentclass[conference]{IEEEtran}
\IEEEoverridecommandlockouts
\usepackage{cite}
\usepackage{amsmath,amssymb,amsfonts}
\usepackage{algorithmic}
\usepackage{algorithm}
\usepackage{graphicx}
\usepackage{textcomp}
\usepackage{xcolor}
\usepackage{booktabs}
\usepackage{array}
\usepackage{hyperref}
\hypersetup{
    colorlinks=true,
    linkcolor=blue,
    citecolor=blue,
    filecolor=magenta,      
    urlcolor=cyan,
    pdftitle={},
}

\begin{document}

% ─────────────────────────────────────────────────────────────────────────────
\title{DAG-Based QoS-Aware Dynamic Task Placement\\
for Networked Multi-Stage Control Pipelines}
%in Cloud--Edge--Robotics Systems}%
% \thanks{Work in progress. A full version with simulation and
% hardware-in-the-loop validation results will be reported in
% a subsequent journal article.}}

\author{
\IEEEauthorblockN{%
  Thien Tran\IEEEauthorrefmark{1},
  Jonathan Kua\IEEEauthorrefmark{1},
  Thuong Hoang\IEEEauthorrefmark{1},
  Minh Tran\IEEEauthorrefmark{2}, 
  Yuemin Ding\IEEEauthorrefmark{3}, and
  Jiong Jin\IEEEauthorrefmark{4}
}
\IEEEauthorblockA{%
    \IEEEauthorrefmark{1}Deakin University, Australia;
    \IEEEauthorrefmark{2}RMIT University, Vietnam;
    \IEEEauthorrefmark{3}University of Navarra, Spain;
}
\IEEEauthorblockA{%
    \IEEEauthorrefmark{4}Swinburne University of Technology, Australia
}
\IEEEauthorblockA{
    \{peter.tran, jonathan.kua, thuong.hoang\}@deakin.edu.au; minh.tranquang@rmit.edu.vn; \\
    yueminding@tecnun.es; jiongjin@swin.edu.au}
}

\maketitle
% Work-in-Progress (WiP)
% ─────────────────────────────────────────────────────────────────────────────
\begin{abstract}
% Final Abstract
Current Physical AI (PAI) relies heavily on closed-loop visual-servoing pipelines, whose perception and planning stages may become computationally intensive onboard due to complex models embedded on robots. 
In practice, offloading the perception task to on-site edges statically is inappropriate for latency-sensitive, precise industrial settings over a standardized industrial network. 
This emphasizes the importance of \emph{Control--Communication--Computing (3C) co-design} in industrial automation: monolithic local execution saturates AI-accelerated machine and robot hardware, while static edge offloading exposes the control loop to network jitter.
Existing adaptive task placement (ATP) controllers can partially address the gap by relocating a single pipeline stage on binary threshold rules, without a multi-stage model and an explicit cost on placement switching. 
In this paper, we propose a directed acyclic graph (DAG) based quality-of-service (QoS)-aware dynamic task placement (DTP) framework for sensing--perception--planning--control pipelines in networked robotics. 
This pipeline is formalized as a DAG with task-level and node-level attributes for compute cost, communication delay, and feasible placement sets; over a small interpretable candidate set (fully local, static offload, hybrid), a window-based cost function combines tail end-to-end latency, deadline violation rate, hardware utilization, and a Hamming-distance switching penalty, and a DTP algorithm with hysteresis and a minimum dwell-time bounds placement chatter.
Our work presents the theoretical framework, a structured qualitative analysis, and a two-phase simulation plus hardware-in-the-loop validation roadmap.
\end{abstract}


% ─────────────────────────────────────────────────────────────────────────────
\section{Introduction}\label{sec1}
% ─────────────────────────────────────────────────────────────────────────────
The compute envelope of factory robotics is being reshaped by Physical AI (PAI) workloads. 
World Foundation Models (WFMs) for surrounding cognitive and agentic visual servoing demand inference costs that exceed the on-board envelope of typical industrial robots, while LLM-based task dispatchers exhibit cold-start latencies of sub-20 seconds, orders of magnitude beyond the sub-millisecond deadlines typical of industrial contact-rich control loops within Vision-Language-Action (VLA) models embedded on PAI~\cite {li26roboclaw, li26opengo}. This mismatch demands a unified networked orchestration framework that binds heterogeneous AI capabilities to the robot–edge compute fabric within the closed-loop timing budget. 

The Cloud-Fog Automation (CFA) reference architecture~\cite{jin_cfa_tii24, jin_cfa_jsac25} articulates a unified vision for networked industrial collaboration across the cloud–fog continuum, while deterministic substrates such as Time-Sensitive Networking (TSN) and 5G-URLLC provide the underlying real-time fabric~\cite{zhang24tsn, ladegourdie2022performance}. Together, they make co-located edge servers a viable extension of the PAI compute envelope, provided placement decisions respect the closed-loop timing budget. However, the layer specifies the runtime decision rule: under workload and network variability that no design-time profile can predict, the deployment question shifts from \lq\lq \textit{whether}\rq\rq~to offload to \lq\lq \textit{which}\rq\rq~pipeline stages run, where by being decided online~\cite{tran_usv_indin25, tran_cftel_indin25}.

A single static placement is inadequate for this regime due to the non-stationary nature of both compute demand and link delays. These factors vary constantly depending on scene complexity and model branching. Additionally, the cost of a poor placement decision is asymmetric: missing a deadline propagates into actuation jitter, which can violate strict safety envelopes.
Empirical studies on industrial testbeds report deadline-violation rates above 40\% under monolithic local deployment of visual-servoing workloads~\cite{iotvr_jamt_25, tran_dt_jii25}. 
Adaptive placement is, therefore, a deployment necessity; the open question is what such a rule should observe, what it should optimize, and how it should be stabilized for deployment.

Directed acyclic graph (DAG) based task offloading has been studied extensively in mobile edge computing (MEC)~\cite{tong26dagmec, liu2023dependent, afrin19dag}, but these models target \lq\lq batch workflows\rq\rq~and optimize average latency or energy rather than deadline violation rate ($V_D$) under closed-loop control. 
Our previous adaptive task placement (ATP) on robot testbeds demonstrates the value of dynamic offloading~\cite{nguyen2025task}. 
However, each system relocates only a \lq\lq single\rq\rq~pipeline stage using binary threshold rules on individual fixed metrics. Therefore, it provides no formal DAG model, no multi-objective cost function, and no analytical resistance to task placement oscillation.

\begin{figure*}[t]
  \centering
  \includegraphics[width=\textwidth]{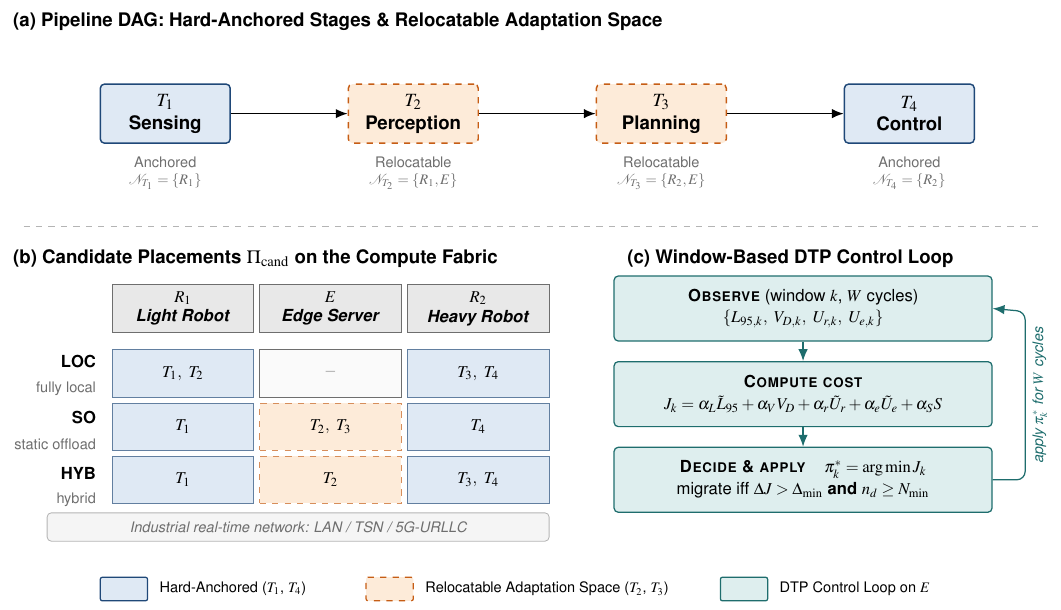}
  \caption{DAG-QoS DTP system architecture.
    \textbf{(a)}~The four-stage sensing--perception--planning--control
    pipeline as a DAG: $T_1$, $T_4$ are hard-anchored;
    $T_2$, $T_3$ form the relocatable adaptation space.
    \textbf{(b)}~Three candidate placements (LOC, SO, HYB) on the
    $R_1$/$E$/$R_2$ compute fabric.
    \textbf{(c)}~Window-based DTP loop on $E$: per-window QoS
    observation, cost $J_k$, and decision $\pi^*_k$ with hysteresis
    $\Delta_\text{min}$ and dwell-time $N_\text{min}$, maintaining
    $V_D \leq 5\%$.}
  \label{fig:arch}
\end{figure*}

In this paper, we propose a DAG-based QoS-aware DTP framework, the formal multi-objective extension of the empirically validated edge-based QoS-aware ATP controller from previous work.
The system architectural framework is shown in Fig.~\ref{fig:arch}. We present the complete theoretical framework, DAG model, cost function, and DTP algorithm, with the structured qualitative analysis and the validation roadmap. 
Our work contributes the DAG pipeline model that distinguishes hard-anchored sensing and control stages from the relocatable perception and planning adaptation space, agnostic to the LAN/TSN/5G-URLLC substrate. 
On this model, we define the multi-objective QoS cost function and the window-based DTP algorithm with hysteresis and minimum dwell-time that bounds placement chatter, formalizing the 3C principles within the single~per-window~optimization.

The remainder of this paper is organized as follows. Section~\ref{sec2} presents the background for this paper and surveys related work, Section~\ref{sec3} presents the DAG pipeline model for multi-stage control pipelines and develops the multi-objective QoS cost and the window-based DTP algorithm, and Section~\ref{sec4} discusses research findings and presents a roadmap for future research directions. Section~\ref{sec5} concludes the paper.
% =============================================================================
\section{Background and Related Work}\label{sec2}
% \noindent\textbf{DAG-based task offloading in MEC.} 
Modeling an application as a DAG and offloading its tasks across tiers is a mature line of work in MEC~\cite{liu2023dependent}. 
Recent efforts target inter-task dependency and delay sensitivity, for example, distributed offloading of dependent, delay-sensitive tasks across multi-operator networks~\cite{asheralieva2024dependent} and federated deep-reinforcement-learning schedulers with automated hyperparameter tuning~\cite{tong2024dagmec}. 
These formulations, however, optimize batch-workflow objectives such as average makespan or energy under one-shot job arrivals. 
None targets the per-cycle ($V_D$) of a periodic, closed-loop control pipeline, nor prices the cost of re-deciding a placement while the loop is running, the two properties that dominate stability in networked control. 

% \noindent\textbf{Cloud, fog, and edge robotics.}
On physical robot platforms, adaptive offloading has been advanced by ROS-native frameworks, most prominently the FogROS2 lineage~\cite{ichnowski2023fogros2}. 
FogROS2-LS adds latency-sensitive, location-independent service selection, dynamically switching among replicated cloud/edge deployments to meet a latency target~\cite{chen2024fogros2ls}. 
Such systems nonetheless treat placement as a binary, single-service decision and report end-to-end latency rather than a control-grade deadline metric; they do not model a multi-stage sensing--perception--planning--control pipeline as a scheduling object.
Our QoS-aware ATP controller empirically validates dynamic relocation of a single perception stage on a two-robot-plus-edge testbed, but (i)~relocates only one stage; (ii)~decides via binary thresholds on individual metrics; and (iii)~carries no explicit switching-cost model to bound placement chatter.

% \noindent\textbf{Physical AI and the runtime deployment gap.}
A growing body of Physical AI work indicates that the barrier to deploying VLA models on robots is increasingly the runtime (real-time inference under tight compute, energy, and latency budgets) rather than model capability alone~\cite{sapkota2026vla, ma_vla_tnnls26}. 
Survey and hardware-characterization studies report VLA inference latencies incompatible with the sub-second control cycles of contact-rich manipulation and identify on-robot deployment, not benchmark accuracy~\cite{zhou_xpu_ar26}. 
This literature motivates an orchestration layer that decides \emph{where} such workloads execute, subject to an explicit closed-loop timing budget, to improve the overall system load balance.

% \noindent\textbf{Positioning.}
In this work, we formalize the multi-stage, multi-objective, stability-aware DTP framework for sensing--perception--planning--control pipelines: it separates hard-anchored from relocatable stages via a DAG model, optimizes a windowed QoS cost dominated by the deadline-violation rate, and bounds placement oscillation with a Hamming-distance switching penalty under hysteresis and minimum-dwell-time constraints. 

% \newpage

\section{DAG-based QoS Dynamic Task Placement}\label{sec3}
% ─────────────────────────────────────────────────────────────────────────────
\subsection{Theoretical System and DAG-based Model}
% ─────────────────────────────────────────────────────────────────────────────
\subsubsection{Compute Nodes and Workload DAG}
Let $\mathcal{N} = \mathcal{R} \cup \mathcal{E} \cup \mathcal{C}$ denote the global compute node set, comprising robots~--~$\mathcal{R}$, edge servers~--~$\mathcal{E}$, and (optionally or extensively) cloud nodes~--~$\mathcal{C}$.
The target empirical factory configuration is $R_1$ (light robot with monocular camera), $R_2$ (heavy robot driving a manipulator), and $E$ (edge server), interconnected by an industrial real-time network with configurable delay and jitter. The framework is agnostic to the link technology, engineered LAN, TSN, or 5G-URLLC~\cite{zhang24tsn}, and is parameterized by the per-edge delay distribution $d_{ij}(\cdot)$.

A control pipeline is modeled as a directed acyclic graph~--~$G = (\mathcal{V}, \mathcal{E}_G)$, with task set $\mathcal{V} = \{T_1, T_2, T_3, T_4\}$ corresponding to four stages, and linear precedence $\mathcal{E}_G = \{(T_1, T_2), (T_2, T_3), (T_3, T_4)\}$. Each task $v \in \mathcal{V}$ carries:
\begin{itemize}
  \item The feasible placement set $\mathcal{N}_v \subseteq \mathcal{N}$;
  \item The compute-time function
    $c_v: \mathcal{N}_v \to \mathbb{R}_{>0}$; and
  \item The CPU-utilization contribution
    $u_v: \mathcal{N}_v \to [0, 1]$.
\end{itemize}
Denote that each precedence edge $(v_i, v_j) \in \mathcal{E}_G$ incurs a communication delay $d_{ij}(n_i, n_j) \geq 0$ depending on link characteristics and payload size of the associated operations.

\subsubsection{Placement Mapping and Architectural Invariant}

A placement mapping $\pi: \mathcal{V} \to \mathcal{N}$ assigns each task to a feasible node. The visual servoing pipeline imposes:
\begin{align}
  \mathcal{N}_{T_1} &= \{R_1\}, &
  \mathcal{N}_{T_4} &= \{R_2\}, \label{eq:anchors}\\
  \mathcal{N}_{T_2} &= \{R_1, E\}, &
  \mathcal{N}_{T_3} &= \{R_2, E\}. \label{eq:relocatable}
\end{align}
This encodes the key architectural invariant of the multi-PAI System in agentic industrial automation operation:
\begin{itemize}
  \item \textit{Hard-anchored stages} ($T_1$, $T_4$) remain on physical AI-accelerated machines and robots, preserving sub-millisecond local operations regardless of network state.
  \item \textit{Relocatable adaptation space} ($T_2$, $T_3$) migrates across the Cloud--Edge--Robotics continuum (native edge tiers) subject to the QoS cost function.
\end{itemize}

\subsubsection{Control Loop and End-to-End (E2E) Latency}

The control loop is periodic with period $P$ and deadline $D \leq P$. Under placement $\pi$, the nominal E2E latency is:
\begin{equation}
  L(\pi) \approx \sum_{v\in\mathcal{V}} c_v(\pi(v))
        + \!\!\sum_{(v_i,v_j)\in\mathcal{E}_G}\!\!
          d_{ij}(\pi(v_i), \pi(v_j)).
  \label{eq:latency}
\end{equation}
Denote that $L$ is stochastic in practice due to workload variability and network jitter during operations; high-percentile values therefore serve as the primary QoS-aware target.

% ─────────────────────────────────────────────────────────────────────────────
\subsection{QoS-Aware Cost Model}
% ─────────────────────────────────────────────────────────────────────────────
\subsubsection{Per-Window QoS Metrics}

The native DTP algorithm operates over observation windows of $W$ control cycles indexed by $k$. Per single window, the original four QoS metrics are collected as follows:
\begin{itemize}
  \item $L_{95,k}$: Empirical 95th-percentile E2E latency, capturing tail behavior from network outliers and bursty compute;
  \item $V_{D,k} \in [0,1]$: Deadline violation rate (primary SLA);
  \item $U_{r,k}$, $U_{e,k}$: CPU utilization across robot/edge nodes.
\end{itemize}
Latency and utilization are normalized by design targets:
\begin{equation}
  \tilde{L}_{95,k} = \frac{L_{95,k}}{\bar{L}}, \quad
  \tilde{U}_{r,k}  = \frac{U_{r,k}}{\bar{U}_r}, \quad
  \tilde{U}_{e,k}  = \frac{U_{e,k}}{\bar{U}_e}.
  \label{eq:normalise}
\end{equation}
$V_{D,k}$ is bounded in $[0,1]$ and requires no normalization.

\subsubsection{Switching Penalty and Cost Function}
To penalize frequent migration, a switching penalty based on the Hamming distance between placements is applied:
\begin{equation}
  S(\pi_k, \pi_{k-1}) = \frac{1}{|\mathcal{V}|}
    \sum_{v\in\mathcal{V}}
    \mathbf{1}\bigl\{\pi_k(v) \neq \pi_{k-1}(v)\bigr\}.
  \label{eq:hamming}
\end{equation}
This~term~suppresses~\lq\lq placement chatter\rq\rq,~high-frequency oscillation that degrades network stability and is the infrastructure-level manifestation of trust latency~\cite{andreoli25tl}. Given non-negative weights, the QoS-aware cost for window $k$ is:
\begin{equation}
  J_k = \alpha_L \tilde{L}_{95,k}
      + \alpha_V V_{D,k}
      + \alpha_r \tilde{U}_{r,k}
      + \alpha_e \tilde{U}_{e,k}
      + \alpha_S S(\pi_k, \pi_{k-1}).
  \label{eq:cost}
\end{equation}
The tension between $\alpha_V V_{D,k}$ (deadline aggression) and $\alpha_S S(\pi_k,\pi_{k-1})$ (migration resistance) formalizes the core control--infrastructure trade-off. In industrial settings, $\alpha_V > \alpha_L > \alpha_S$ reflects prioritizing SLA compliance as the primary, raw latency as the secondary, and stability as a soft constraint.

\subsubsection{Constrained Placement Problem}
To keep per-window evaluation tractable on edge hardware, we originally restrict to a small interpretable candidate set $\Pi_\text{cand}$ comprising the following three pilot placements:
\begin{itemize}
  \item \textbf{LOC} (fully local): $T_2$ on $R_1$, $T_3$ on $R_2$;
  \item \textbf{SO} (static offload): $T_2$ and $T_3$ on $E$;
  \item \textbf{HYB} (hybrid): $T_2$ on $E$, $T_3$ on $R_2$.
\end{itemize}
The window-level placement problem is
\begin{equation}
  \pi^*_k \in \arg\min_{\pi \in \Pi_\text{cand}}
    J_k(\pi; \pi_{k-1}),
  \label{eq:opt}
\end{equation}
subject to $L_{95,k}(\pi) \leq L_\text{max}$ and per-node utilisation
caps $U_{n,k}(\pi) \leq U_\text{max}$ for all $n \in \mathcal{N}$.

% ─────────────────────────────────────────────────────────────────────────────
\subsection{Dynamic Task Placement Algorithm}
% ─────────────────────────────────────────────────────────────────────────────

\subsubsection{Window-Based DTP Procedure}

Algorithm~\ref{alg:dtp} describes the DTP procedure running on the edge node~--~$E$. Two parameters jointly enforce hysteresis:
\begin{itemize}
  \item $N_\text{min}$: Minimum dwell windows; prevents rapid flapping when metrics fluctuate near thresholds;
  \item $\Delta_\text{min}$: Cost-improvement threshold; ensures only statistically significant improvements trigger migration.
\end{itemize}
Denote that per-window complexity is $O(|\Pi_\text{cand}|)$, which is always lightweight on any edge node if task placements occur.

\begin{algorithm}[t]
\caption{Window-Based Dynamic Task Placement (DTP)}
\label{alg:dtp}
\begin{algorithmic}[1]
\REQUIRE $W$;\; $(\alpha_L,\alpha_V,\alpha_r,\alpha_e,\alpha_S)$;\;
  $\Delta_\text{min}$;\; $N_\text{min}$;\; $\Pi_\text{cand}$
\STATE Init $\pi_0 \in \Pi_\text{cand}$;\quad dwell $n_d \leftarrow 0$
\FOR{$k = 1, 2, \ldots$}
  \STATE Apply $\pi_{k-1}$ for $W$ cycles;
    collect $\{L_{95,k}, V_{D,k}, U_{r,k}, U_{e,k}\}$
  \STATE Compute $J_k(\pi_{k-1}; \pi_{k-2})$ via \eqref{eq:cost}
  \IF{$n_d < N_\text{min}$}
    \STATE $n_d \leftarrow n_d{+}1$;\; $\pi_k \leftarrow \pi_{k-1}$;\;
      \textbf{continue} \hfill\COMMENT{enforce min.\ dwell}
  \ENDIF
  \STATE Estimate $J_k(\pi; \pi_{k-1})$ for all
    $\pi \in \Pi_\text{cand}$
  \STATE $\pi^*_k \leftarrow \arg\min_\pi\, J_k(\pi; \pi_{k-1})$
  \STATE $\Delta J \leftarrow
    J_k(\pi_{k-1}; \pi_{k-2}) - J_k(\pi^*_k; \pi_{k-1})$
  \IF{$\Delta J > \Delta_\text{min}$}
    \STATE $\pi_k \leftarrow \pi^*_k$;\; $n_d \leftarrow 0$
    \hfill\COMMENT{migrate}
  \ELSE
    \STATE $\pi_k \leftarrow \pi_{k-1}$;\; $n_d \leftarrow n_d{+}1$
    \hfill\COMMENT{hold}
  \ENDIF
\ENDFOR
\end{algorithmic}
\end{algorithm}

\subsubsection{Estimating QoS for Non-Active Placements}

Scoring placements not currently active requires estimating their idle QoS metrics. The three initial complementary mechanisms are supported along the operation loop as follows:
\begin{itemize}
  \item \textbf{Static profiling:} Compute times $c_v(n)$ and communication delays measured offline and stored as lookup tables.
  \item \textbf{Online emulation:} A small fraction of cycles is routed through each alternative in the background to update estimates continuously during the non-placement stage.
    \item \textbf{Conservative scaling:} Upper-bound estimates derived from current measurements or passively monitored feedback and known compute/network ratios.
\end{itemize}
The original algorithm is agnostic to the choice, provided cost comparisons remain qualitatively consistent for DTP.

\section{Discussion and Roadmap}\label{sec4}
% ─────────────────────────────────────────────────────────────────────────────
\subsection{Qualitative Analysis and Deployment Context}
% ─────────────────────────────────────────────────────────────────────────────
\subsubsection{Anticipated Placement Trajectories}
Table~\ref{tab:scenarios} summarizes the expected steady-state DTP behaviors relative to LOC and SO baselines across four pilot canonical stress scenarios. 
The policy converges toward LOC under adverse network conditions and toward SO under robot CPU stress; HYB dominates when compute and communication resources are jointly constrained, by offloading the heavier $T_2$~(Perception) to $E$ while keeping $T_3$~(Planning) on $R_2$.

% \begin{table}[t]
% \caption{Expected dominant DTP placement per stress scenario.}
% \label{tab:scenarios}
% \begin{center}
% \renewcommand{\arraystretch}{1.18}
% \begin{tabular}{|p{1.45cm}|p{1.45cm}|p{1.45cm}|p{1.65cm}|}
% \hline
% \textbf{Scenario} & \textbf{LOC} & \textbf{SO} &
% \textbf{DTP (expected)} \\
% \hline
% Baseline &
%   Low $V_D$;\par high $U_r$ &
%   Low $U_r$;\par moderate $V_D$ &
%   Mostly LOC;\par occasional HYB \\
% \hline
% Robot CPU stress &
%   High $V_D$\par ($U_r$ sat.) &
%   Mitigated;\par higher $L_{95}$ &
%   Converges to SO \\
% \hline
% Edge CPU stress &
%   Stable;\par $U_r$ elevated &
%   High $V_D$\par ($U_e$ sat.) &
%   LOC or HYB \\
% \hline
% Network impairment &
%   Low\par variance &
%   High $V_D$\par via jitter &
%   Converges to LOC \\
% \hline
% \end{tabular}
% \end{center}
% \end{table}

\subsubsection{Co-Design Implications and the PAI Nervous System}
Cast as the CFA 3C co-design principle-based orchestration layer, the DAG-based QoS-aware DTP framework provides three concrete guarantees against the failure modes intrinsic to networked factory control following industrial standards:

\begin{itemize}
  \item The hard-anchoring of $T_1$ and $T_4$ preserves the sub-millisecond reflex loop on local hardware, decoupling actuation safety and dependencies from network state;
  \item The relocatable adaptation space accommodates the heterogeneous compute demands of $T_2$ and $T_3$, including emerging VLA/LLM workloads, without overloading machine hardware or hard-coding offload decisions;
  \item The Hamming-distance switching penalty in \eqref{eq:hamming} provides a mathematically bounded guarantee against placement chatter, translating directly to operator-observable control-loop stability and to the deterministic timing budgets expected of TSN-class substrates~\cite{zhang24tsn}.
\end{itemize}

Technically, a factory automation system that cannot guarantee $V_D \leq 5\%$ across the deployment envelope cannot credibly claim Industry~4.0 production maturity~\cite {andreoli25tl} toward industrial agentic automation for Industry 5.0 readiness.

% ─────────────────────────────────────────────────────────────────────────────
\subsection{Empirical Strategy and Validation Roadmap}
% ─────────────────────────────────────────────────────────────────────────────
Validation proceeds in two main phases, each with concrete parameters drawn directly from the physical robotic testbed to ensure direct comparability and reflection on the previous empirical LOC, SO, and ATP baselines.

\begin{table}[!b]
\caption{Expected dominant DTP placement per stress scenario.}
\label{tab:scenarios}
\begin{center}
\renewcommand{\arraystretch}{1.18}
\begin{tabular}{p{1.45cm}|p{1.45cm}|p{1.45cm}|p{1.65cm}}
% \hline
\textbf{Scenario} & \textbf{LOC} & \textbf{SO} &
\textbf{DTP (expected)} \\
\hline
Baseline &
  Low $V_D$;\par high $U_r$ &
  Low $U_r$;\par moderate $V_D$ &
  Mostly LOC;\par occasional HYB \\
\hline
Robot CPU stress &
  High $V_D$\par ($U_r$ sat.) &
  Mitigated;\par higher $L_{95}$ &
  Converges to SO \\
\hline
Edge CPU stress &
  Stable;\par $U_r$ elevated &
  High $V_D$\par ($U_e$ sat.) &
  LOC or HYB \\
\hline
Network impairment &
  Low\par variance &
  High $V_D$\par via jitter &
  Converges to LOC \\
% \hline
\end{tabular}
\end{center}
\end{table}

\subsubsection{Phase~1 -- Discrete-Event Simulation}
A discrete-event simulator parameterized will validate cost-function sensitivity and hysteresis bounds against predictions. Key parameters are:
\begin{itemize}
  \item Control period $P \in [20, 50]$\,ms with deadline $D \leq P$;
  \item Baseline link RTT of 1--2\,ms representative of an engineered industrial LAN, with TSN-class jitter envelopes as a sensitivity axis following standardization~\cite{zhang24tsn};
  \item CPU stress profiles applied to the $R_1$-equivalent node;
  \item Gaussian network fault injection (e.g., $\mu = 25$\,ms, $\sigma = 5$\,ms, $p_\text{loss} = 2\%$), plus additional extended verification.
\end{itemize}

\subsubsection{Phase~2 -- Hardware-in-the-Loop (HIL)} Phase~2 enables direct quantitative comparison between the proposed multi-stage DTP and the single-stage ATP baseline.
The modular cloud-edge-robotics testbed will provide:

\begin{itemize}
  \item Measured compute profiles $c_v(n)$ on each physical node;
  \item Empirical network delay distributions on the $R_1$--$E$ and $E$--$R_2$ links with extended IIoT devices/PLCs;
  \item Empirical CDFs of $L_{95}$, $V_D$, $U_r$, and $U_e$ for LOC, SO, ATP, and DAG-QoS DTP across all stress scenarios.
\end{itemize}

\subsubsection{Research Directions}
The two-phase plan targets four research questions/directions:
\begin{itemize}
  \item \textbf{RQ1.} Does concurrent $T_2$+$T_3$ placement reduce
    $V_D$ further than single-stage ATP under CPU stress?
  \item \textbf{RQ2.} Does routing $T_3$ to $R_2$ during network
    faults outperform binary $R_1/E$ reversion?
  \item \textbf{RQ3.} Does the Hamming-distance penalty provide
    finer-grained chatter suppression than threshold hysteresis?
  \item \textbf{RQ4.} What is the sensitivity of DTP to the weight
    configuration $(\alpha_L, \alpha_V, \alpha_r, \alpha_e,
    \alpha_S)$ under each scenario?
\end{itemize}

% ─────────────────────────────────────────────────────────────────────────────
\section{Conclusions and Future Work}\label{sec5}
% ─────────────────────────────────────────────────────────────────────────────
% The pipeline is originally formalized as a DAG with task-level and node-level attributes for compute cost, communication delay, and feasible placement sets; over a small interpretable candidate set (fully local, static offload, hybrid), a window-based cost function combines tail end-to-end latency, deadline violation rate, hardware utilization, and a Hamming-distance switching penalty, and a lightweight DTP algorithm with hysteresis and a minimum dwell-time bounds placement chatter under metric noise. 

% The framework extends our prior single-stage QoS-aware ATP controller to concurrent multi-task placement under a substrate-agnostic delay model spanning LAN, TSN, and 5G-URLLC. 

In this paper, we presented the DAG-based QoS-aware DTP, the formal multi-objective extension of our edge-based QoS-aware ATP framework. By formalizing the visual-servoing pipeline as the DAG, introducing the window-based cost function with the Hamming-distance switching penalty, and generalizing the ATP algorithm to concurrent multi-task placement, the extended framework addresses three structural limitations of single-task adaptive placement: limited scope, threshold-based decisions, and the absence of a formal switching cost model. The result is a runtime determinism layer compatible with the CFA-based 3C co-design central to factory communications and mature industrial automation. Beyond factory robotic automation, this middle layer will act as one of the key cores in \lq\lq \textit{PAI's Nervous System}\rq\rq: the runtime infrastructure that enables LLM agents and VLA models to operate reliably under real compute and network variability on different PAIs and nearby edges/clouds, making industrial-grade trustworthiness in Industry~5.0~\cite{jiang2026comprehensive,boateng2025survey}.

% Future research directions must extend DAG-based QoS-aware DTP toward full deployment, but not limited to:
% \begin{itemize}
%   \item Physical ROS~2 deployment on the testbed of~\cite{tran_atp_indin25};
%   \item Joint scheduling with deterministic substrates, including stream reservation and multi-robot admission control;
%   \item Adaptive weight tuning via online Bayesian optimization or lightweight meta-learning for modeling;
%   \item Additional QoS dimensions such as Age-of-Information (AoI) and safety margins.
% \end{itemize}

Future work will extend the proposed framework in this paper towards full deployment through: (i) empirical validation on the physical testbed; (ii) expansion of the candidate placement set beyond LOC/SO/HYB and extension to multi-robot fleets where a single edge serves several DTP loops concurrently; (iii) online cost-weight adaptation of hyperparameters via Bayesian optimization or lightweight meta-learning; (iv) joint compute--network co-scheduling with deterministic substrates, elevating the network from a delay distribution to a co-optimized hardware resource, and considering new QoS dimensions such as Age-of-Information (AoI).

% ─────────────────────────────────────────────────────────────────────────────
%%% References
\bibliographystyle{IEEEtran}
\bibliography{refs}

\end{document}